# Liquid-Fed Pulsed Plasma Thruster for Propelling Nanosatellites

Adam R. Patel, Yunping Zhang, and Alexey Shashurin[*]

School of Aeronautics and Astronautics, Purdue University, West Lafayette, USA

**Abstract**

This letter presents a novel micropropulsion system for nanosatellite applications - a liquid fed pulsed-plasma thruster (LF-PPT) comprised of a Lorentz-force pulsed plasma accelerator (PPA) and a low-energy surface flashover (LESF) igniter. A 3 μF / 2 kV capacitor bank, offering shot energies of < 6 J, supported PPA current pulse durations of ~ 16 μs with observed peaks of 7.42 kA. Plasma jet exhaust velocity was measured at ~ 32 km/s using a time-of-flight technique via a set of double probes located along the jet's path. Intensified charge coupled device (ICCD) photography was concurrently leveraged to visualize plasma dynamics and mechanisms of the ignition / acceleration events. A peak thrust and impulse bit of 5.8 N and 35 μN·s, respectively, were estimated using large-area Langmuir probe measurements of total ion flux produced by the thruster.

**Introduction**

The rapid development and application of nanosatellite technology has vastly accelerated mission complexity – sparking interest in robust, low power, and high specific impulse micropropulsion systems. Pulsed plasma thrusters (PPTs) have been extensively investigated and employed to fill such roles, debuting on the 1964 Soviet Zond 2. Like magnetoplasmadynamic engines, PPTs accelerate plasma propellant through the Lorentz-force - preferably with a minimum of thermal and electromagnetic loss [1]. In lieu of an applied external magnetic field, an induced component resulting from current traversal through the electrodes and plasma supplies the *B*-field required for acceleration. High discharge currents (typically $10^4$-$10^6$ A) enable sufficiently high J x B Lorentz-force magnitudes [2]. Typically, PPTs utilize a capacitor bank for energy storage, which is subsequently converted into kinetic motion, heating, and propellant ionization upon initiation of the discharge. These capacitors can be charged by on-board or solar energy. Accordingly, it can be implemented in power-limited nanosatellites for attitude control maneuvers. Another particularly significant benefit to pulsed operation is increased thrust efficiency, enabled by the ability to operate at higher discharge currents without thermal damage to electrodes. The operation at these higher discharge currents results in a quadratic increase in thrust. In addition, unsteady operation can further prevent energy losses from radiation and electron-ion thermalization [2].

Typical nanosatellite PPT designs are ablative in nature (APPTs) and operate with solid phase propellants such as polytetrafluoroethylene or similar fluorocarbons [3, 4, 5, 6, 7]. The surfaces of these propellants are vaporized with high currents, and the resulting plasma is accelerated to produce thrust. This process places a harsh limit on efficiency (typically <15 %) – plagued by late-ablation and the presence of thermally expelled macroparticles [4]. Impulse bit variability (nonuniform ablation), low mass flow control, and contamination [8] pose further problems to the implementation of APPTs. These disadvantages are often disregarded, however, as high reliability and low tankage fractions may provide an attractive thruster option.

---

[*] Corresponding Author: Alexey Shashurin (ashashur@purdue.edu)

Gas-fed pulsed plasma thrusters (GF-PPTs) pose a stark contrast to their ablative contemporaries. These variants offer high efficiencies of 20 - 70% [1, 9] and precision mass flow-control at the cost of complex injection systems. These apparatuses can strain implementation in satellites which have limited volumetric capabilities, namely CubeSats. Gas injection arrangements also place a mechanical limit on firing frequency due to the presence of mechanical solenoid valves [2]. Further, difficulties often arise in attempting to establish a desired mass density distribution before ignition – where inductive delay and gas injection must be properly timed.

Another set of limiting factors associated with currently available PPTs refers to an igniter subsystem which is imperative to initiate PPT firing events. Alongside carbonization, igniters are often the limiting factor in thruster operational lifetime [10]. A variety of ignition techniques have been used with varying success ranging from 400 to 1,000,000 firing cycles [3, 9, 11]. In our previous work, it has been demonstrated that low-energy surface flashover (LESF) is an attractive candidate for reliable and long-lasting ignition for micropropulsion systems [12, 13]. LESF modifies classic surface flashover between two metal electrodes separated by a dielectric material by limiting the amount of the energy available for each individual flashover event. This subsequently eliminates damage to the flashover assembly. Upwards of 1.5 million breakdown events were demonstrated by the LESF igniter without significant damage to the assembly.

PPTs with solid and gaseous propellants have been employed with limited degrees of success in nanosatellites. A liquid-fed pulsed plasma thruster could potentially overcome several disadvantages associated with traditional PPT devices such as contamination issues, non-uniform propellant consumption (leading to premature thruster failure), and complex/unreliable propellant feeding systems and igniters. In this work, we present a novel liquid-fed pulsed plasma thruster (LF-PPT) micropropulsion system and demonstrate initial measurements of the thruster parameters.

**Experimental Arrangement**

*Vacuum Chamber*: The experiments were conducted in two vacuum facilities with volumes of 0.069 m$^3$ and 0.66 m$^3$, respectively. Chambers were pumped using diffusion pumps to an ultimate vacuum pressure of $< 6 \cdot 10^{-5}$ Torr. A MKS PDR900 vacuum gauge controller with 925 micropirani transducer was used to read pressures in excess of $10^{-3}$ Torr. For pressures below $10^{-3}$ Torr, a Granville-Phillips 330 controller with iridium-filament ion gauge tube was utilized. Each vacuum chamber was equipped with 15 kV and BNC feedthroughs for high voltage LF-PPT connections and diagnostic equipment. Chambers were equipped with viewports to allow visual observation.

*Electromagnetic Accelerator*: The LF-PPT consists of a pulsed plasma accelerator (PPA) portion and an LESF igniter as shown in Figure 1(a). The PPA was formed by a pair of oxygen-free high thermal conductivity (OFHC) copper electrodes in a parallel-plate configuration. Electrode spacing and width were both measured at 1.27 cm, with an accelerating channel length of 6.5 cm. The LESF igniter was formed by one of the PPA electrodes and an additional electrode placed between the thruster rails. Electrode spacing between LESF electrodes was measured at 2.75 mm. A liquid propellant occupied the space between these electrodes, as shown in the insert image depicted in Figure 1(a). Pentaphenyl trimethyl trisiloxane ($C_{33}H_{34}O_2Si_3$) was used as a propellant in this work due to its excellent dielectric properties and low vapor pressure. Conventionally, $C_{33}H_{34}O_2Si_3$ is used as a diffusion pump working fluid. $C_{33}H_{34}O_2Si_3$ has a high molecular weight, viscosity, boiling point (245° C at 0.06 Torr), and flashpoint (243° C). It is characterized by a low surface tension, low vapor pressure (3 x $10^{-10}$ Torr at 25° C), and low reactivity. The capacitance of the LESF assembly with a $C_{33}H_{34}O_2Si_3$ dielectric insert was measured at 26.6 pF. A back insulator was machined of acrylic to prevent propellant leakage and maintain the thruster's structure. The side walls containing the thruster were also manufactured of acrylic. An optional storage and solenoid valve are additionally illustrated in Figure 1(a), acting as a propellant feed mechanism. Experiments presented in this work did not utilize a propellant feed mechanism. Instead, the PPA assembly was oriented vertically and the interelectrode LESF spacing was manually filled with propellant. Relatively short tests were conducted in this work. <200 firings were accomplished with a single propellant filling without reloading.

Electrical schematics of the thruster are outlined in Figure 1(b). The PPA electrodes were connected to a 3 μF capacitor bank, formed by two 1.5 μF / 2 kV KEMET nonpolar polypropylene capacitors connected in parallel. The capacitor bank was charged to 1.8 kV using a Bertan 225-210-05R high voltage source. The LESF igniter shared the negative PPA electrode as a common, accompanying the high-voltage LESF anode. High voltage up to ~ 8 kV was supplied to the LESF anode through a 500 kOhm current-limiting resistor by means of a Bertan 225-20R high voltage source.

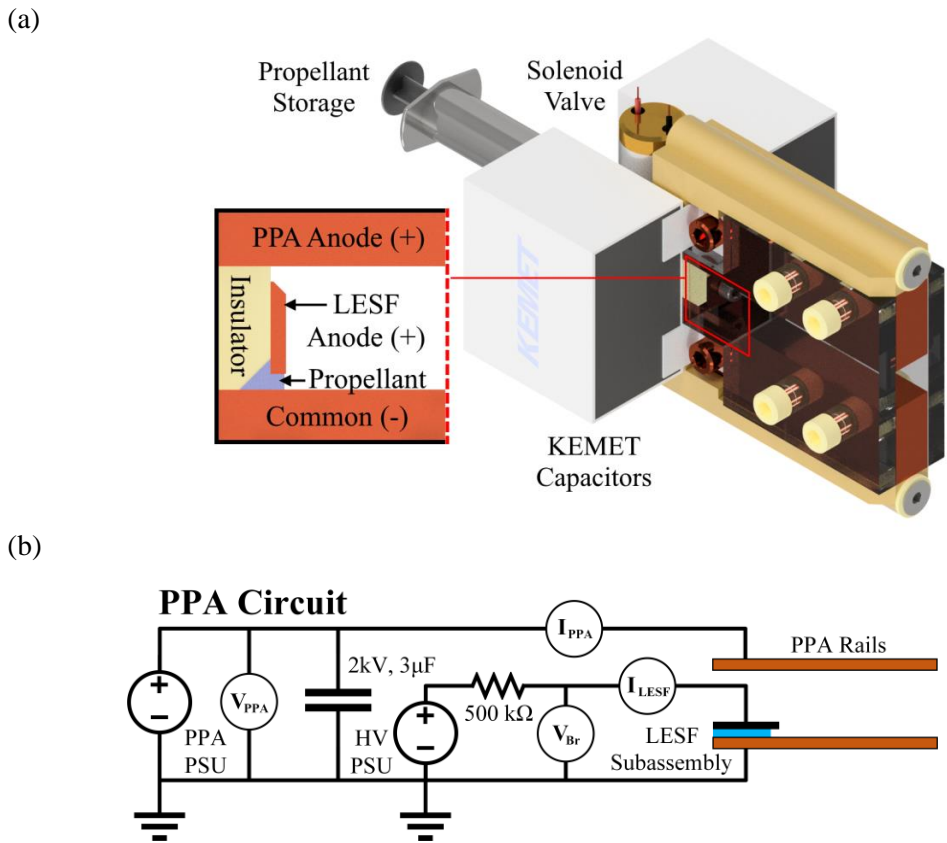

**Figure 1: Liquid-fed pulsed plasma thruster (LF-PPT) with low energy surface flashover (LESF) igniter. (a) CAD rendering. (b) Electrical circuitry of the LF-PPT**.

*Experimental Diagnostics*: A photograph of the LF-PPT equipped with diagnostics is shown in Figure 2. A Pearson 2100 current monitor and Tektronix P6015 voltage probe were utilized to measure the current ($I_{LESF}$) and voltage ($V_{LESF}$) of the LESF igniter. Concurrently, PPA current ($I_{PPA}$) and voltage ($V_{PPA}$) were measured by a Pearson 101 current monitor and Tektronix P2221 probe, respectively. A first order Savitzky-Golay finite impulse response (FIR) digital filter was employed to smooth the discharge current waveforms. In the experimental assembly, a PTFE-insulated 316 stainless-steel extension was utilized to connect the capacitor bank to the PPA, allowing space at the PPA anode for a current monitor (measuring $I_{PPA}$) at the cost of increased impedance. All instrumentation was read through a Teledyne Lecroy WavePro 735Zi 3.5 GHz oscilloscope.

To visualize LESF igniter breakdown and PPA plasma dynamics, a Princeton Instruments PI-MAX4 intensified charge coupled device (ICCD) and Lightfield software were utilized. Long exposure photos were taken by a Nikon d810 camera.

For exhaust velocity determination, a set of three double probes was utilized as shown in Figure 2(a). The probes were located at 5.2, 6.5, and 7.8 cm from the PPA exhaust. Each probe was constructed with a dual channel alumina insulator, 22ga copper wire, and was voltage-biased to 18 V. The circuits for these probes are illustrated in Figure 2(b). Currents $I_{DP1}$-$I_{DP3}$ corresponding to the double probes were measured with three Bergoz FCT-028-0.5-WB current monitors. A first order Savitzky-Golay finite impulse response (FIR) digital filter was employed to smooth the double probes' current waveforms. Note, double probes were used in these measurements rather than single probes in order to eliminate any electrical contact with the thruster electrode assembly which causes generation of noise simultaneously with discharge initiation and problems of identifying signal associated with plasma arrival.

The total ion current generated by the LF-PPT was measured using a large-area plane Langmuir probe (or unguarded Faraday probe) with a diameter of 16.5 cm [14, 15, 16, 17]. The current collected by the probe was directly measured by a Bergoz fast current transformer as shown in Figure 2(b). A shunt resistor of 0.1 Ohm and 36 V lead-acid batteries were utilized in the test circuit. To confirm that EMI contribution to the probe signal is negligible, a series of test experiments with a PTFE-film covered Langmuir probe were conducted prior to taking the probe measurements.

(a)

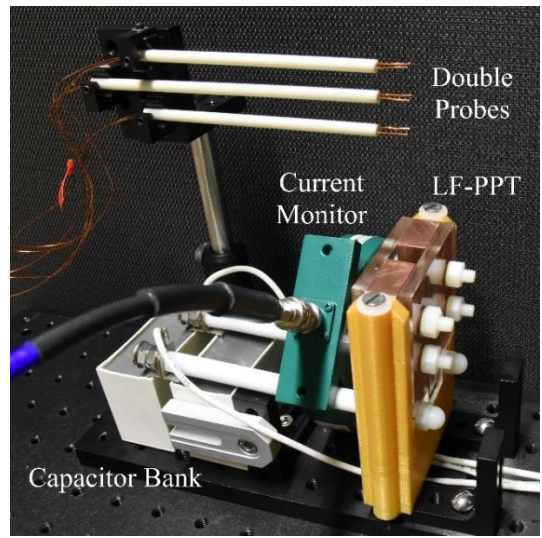

(b)

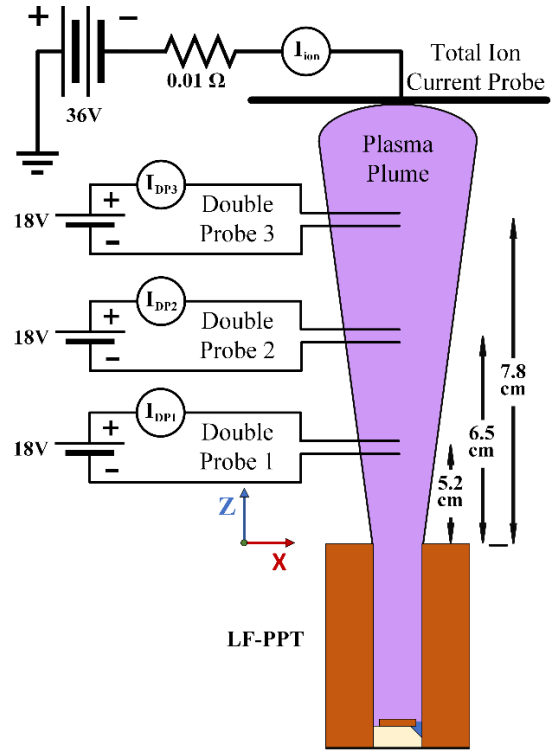

**Figure 2:** LF-PPT assembly equipped with plasma diagnostics of thruster and plume parameters. (a) Photograph of the assembly with installed double probes. (b) Schematics of electrical circuitry used for exhaust plume measurements and overall geometry of the system.

**Experimental Results and Discussions**

*Low Energy Surface Flashover Igniter: V-I* waveforms and a corresponding series of fast photographs of an independent LESF are shown in Figure 3 ($V_{PPA} = 0$ V in this experiment). It was observed that breakdown occurred when $V_{LESF}$ reached about 8 kV. The inception of surface flashover is signified by a drop in the voltage at $t = 0$ as illustrated in Figure 3(a). A subsequent generation of seed plasma shorts the LESF assembly electrodes and triggers a resonant ringing in the LC-circuit, formed by the plasma column shorted flashover assembly. The current oscillations peaked at 12 A and decayed on the time scale of 100-200 ns (this decay time provides an estimate of flashover duration time). The flashover decay time of 100-200 ns is further confirmed by fast ICCD photography taken at moments of time $\tau_1$-$\tau_4$ (35, 65, 100 and 155 ns, respectively) with 3 ns exposure, depicted in Figure 3(b). Note, the ignition events photographed with ICCD were well repeatable, so images shown in Figure 3(b) were acquired in different ignition events. One can see that the flashover plasma vanished at about $t = 100$-155 ns after the flashover initiation. Initial energy stored in the flashover assembly's capacitance of $C = 26.6$ pF prior the breakdown was approximately $E_0 = \frac{1}{2}CV^2 = 0.85$ mJ. The oscillation period of $T=25$ ns observed experimentally was consistent with the shorted LESF assembly inductance of $L=0.6$ μH. Additional details on estimating LESF parameters can be found in our previous works [12, 13].

(a)  (b)

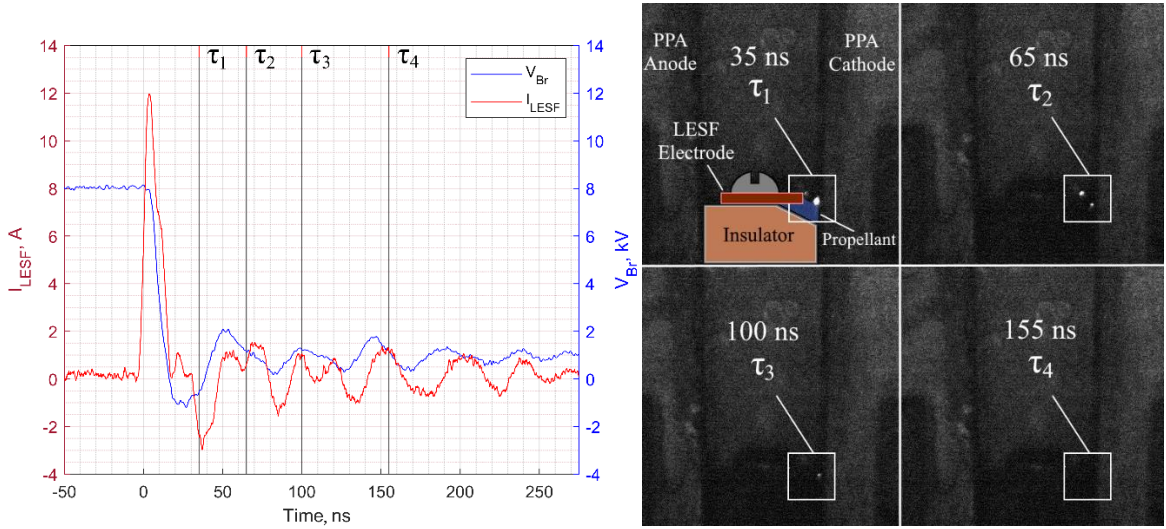

**Figure 3: Simultaneous measurements of electrical parameters and intensified charge-coupled photographs of the LESF ignition event. (a) Current and voltage waveforms during a single flashover event, where timestamps $\tau_1$-$\tau_4$ correspond to Figure 3(b). (b) Photographs of the LESF event taken by an ICCD camera at timestamps $\tau_1$-$\tau_4$ with exposure time 3 ns.**

*Accelerating Channel Dynamics:* It was observed in the experiment that the LESF flashover event triggered PPA discharge when DC voltage was applied to the PPA electrodes. This is illustrated in Figure 4, where voltage and current waveforms of the PPA are presented along with visual ICCD camera observation. Considerable noise was observed around $t = 0$ due to the LESF ignition event. Following the generation of seed plasma by LESF, capacitor voltage ($V_{PPA}$) dropped from an initial value of 1.8 kV and arc current ($I_{PPA}$) rose to a peak of 7.42 kA. The decaying oscillations of current and voltage shifted approximately 90 degrees, indicating operation in the underdamped LCR regime [2, 18]. To visualize the dynamics of plasma acceleration in PPA, ICCD images (100 ns exposure) taken at timestamps $t_1$-$t_5$ of the discharge cycle are illustrated in Figure 4(b). Note, the PPA discharge events photographed with ICCD were well repeatable, so images shown in Figure 4(b) were acquired in different discharge events. A 1 second long exposure photograph of the PPA firing event is additionally depicted for reference. One can clearly see plasma front propagation inside the PPA assembly from the location of seed plasma creation (near the LESF at the bottom of the channel) towards the exit of the PPA channel. Plasma front propagation speed varied in the range 10-30 km/s.

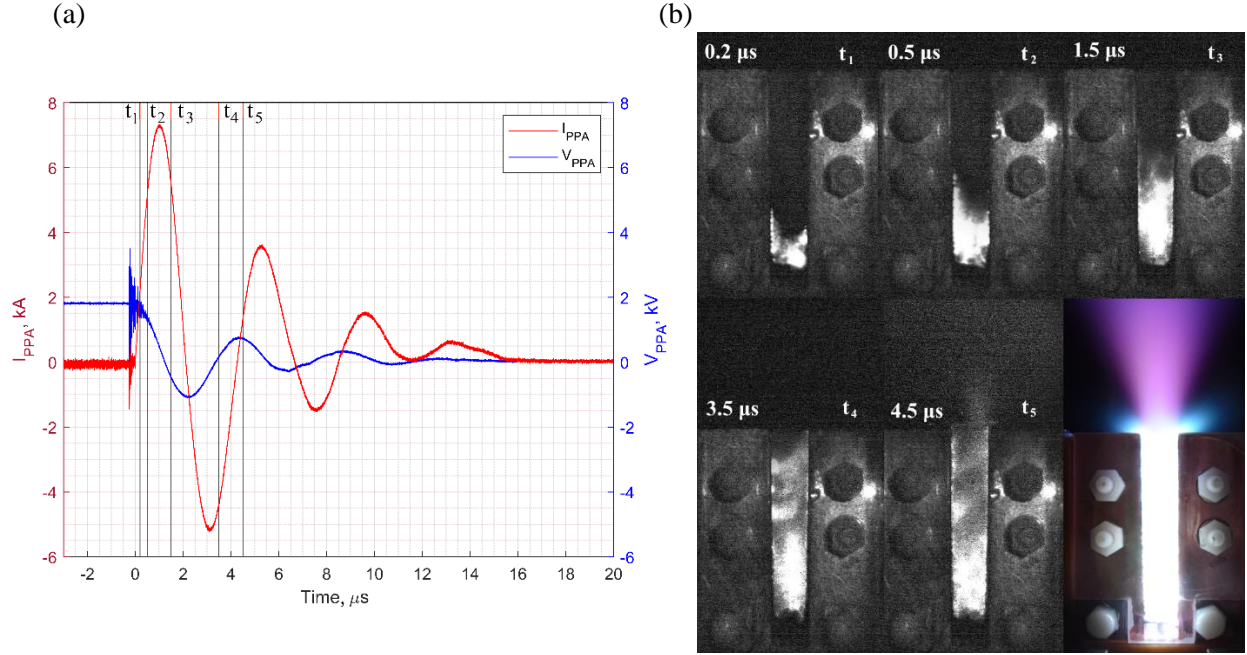

**Figure 4: Simultaneous *V-I* waveform measurements and intensified charge-coupled / long exposure photographs taken during LF-PPT firing. (a) Voltage and current waveforms during a single pulsed-discharge, where timestamps *t₁-t₅* correspond to Figure 4(b). (b) Photographs of the firing event taken by an ICCD camera at timestamps t₁-t₅ with exposure time 100ns. Long exposure (1 s) of the LF-PPT additionally captured for reference.**

*Exhaust Plume Propagation:* The set of three double probes exposed to the LF-PPT exhaust plume is photographed in Figure 5(a). The $I_{PPA}$ and current waveforms measured by the three double probes ($I_{p1}$, $I_{p2}$, and $I_{p3}$), located at $z$ = 5.2, 6.5, and 7.8 cm, are depicted in Figure 5(b). The moment of plasma arrival to each double probe is indicated by a corresponding rise of the double probe current. One can see that plasma reached the probes at $z$ = 5.2, 6.5, and 7.8 cm at 3.9, 4.3, and 4.7μs, respectively. Plasma arrival times were used in conjunction with probe spacing to estimate exhaust velocity at $v_i \approx$ 32 km/s.

(a) (b)

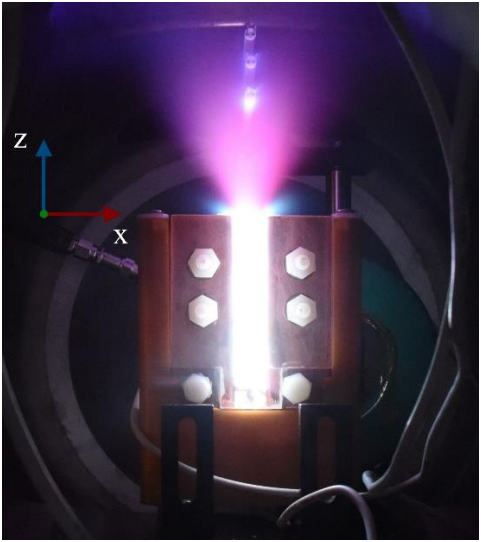 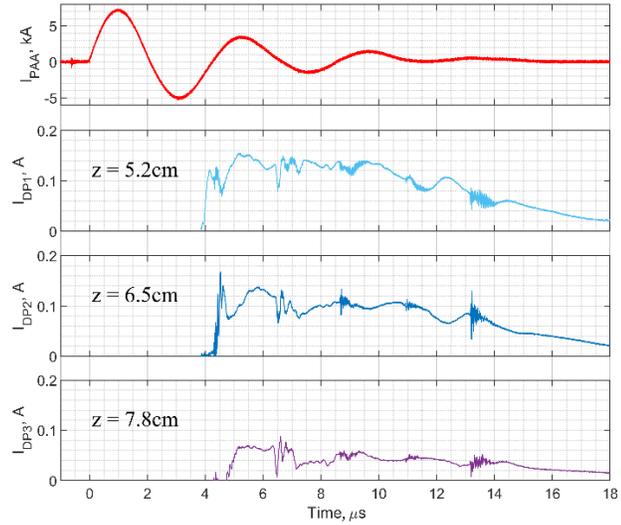

**Figure 5: (a) Double probes submurged in the LF-PPT plume. Probes I$_{DP1}$-I$_{DP3}$ are spaced 5.2, 6.5, and 7.8cm from the thruster exit, respectively. (b) Time-of-flight measurements of LF-PPT exhaust velocity, where I$_{DP1}$-I$_{DP3}$ correspond to current measured in 18V biased double probes with Rogowski coils.**

*Total Current Measurement:* Total ion current produced by the PPA ($I_{ion}$) and measured by the large-area Langmuir probe is presented in Figure 6, along with the $I_{PPA}$ waveform. The negative current precursor seen at *t*~7 μs after PPA discharge ignition can be potentially attributed to the arrival of the front of fast ions or electrons [19]. The bulk of plasma arrival is indicated by the positive pulse of ion current which peaked at *t*~20 μs. Instantaneous thrust produced by the plasma ions can be roughly estimated from the measured total ion current through $T = \dot{m}_i v_i = \frac{I_{ion}}{Z|e|} M v_i$, where $\dot{m}_i$, $v_i$, $Z$ and $M$ are the ion mass flow rate, average exhaust velocity, average ion charge number, and propellant molecular mass, respectively. Using measured ion velocity $v_i$~32 km/s, and assuming $Z = 1$ and propellant ion mass to be 546.9 amu, one can estimate peak thrust value on the order of $T \approx 5.8$ N corresponding to the peak ion current $I_{ion} = 32$ A. Impulse bit ($\Delta P = \int T dt$) of the LF-PPT can be estimated at 35 μN·s using a simple trapezoidal approximation. Note, decomposition of the propellant molecule in PPA discharge is possible, so 5.8 N peak thrust and 35 μN·s impulse bit provide the upper estimate for the corresponding parameters.

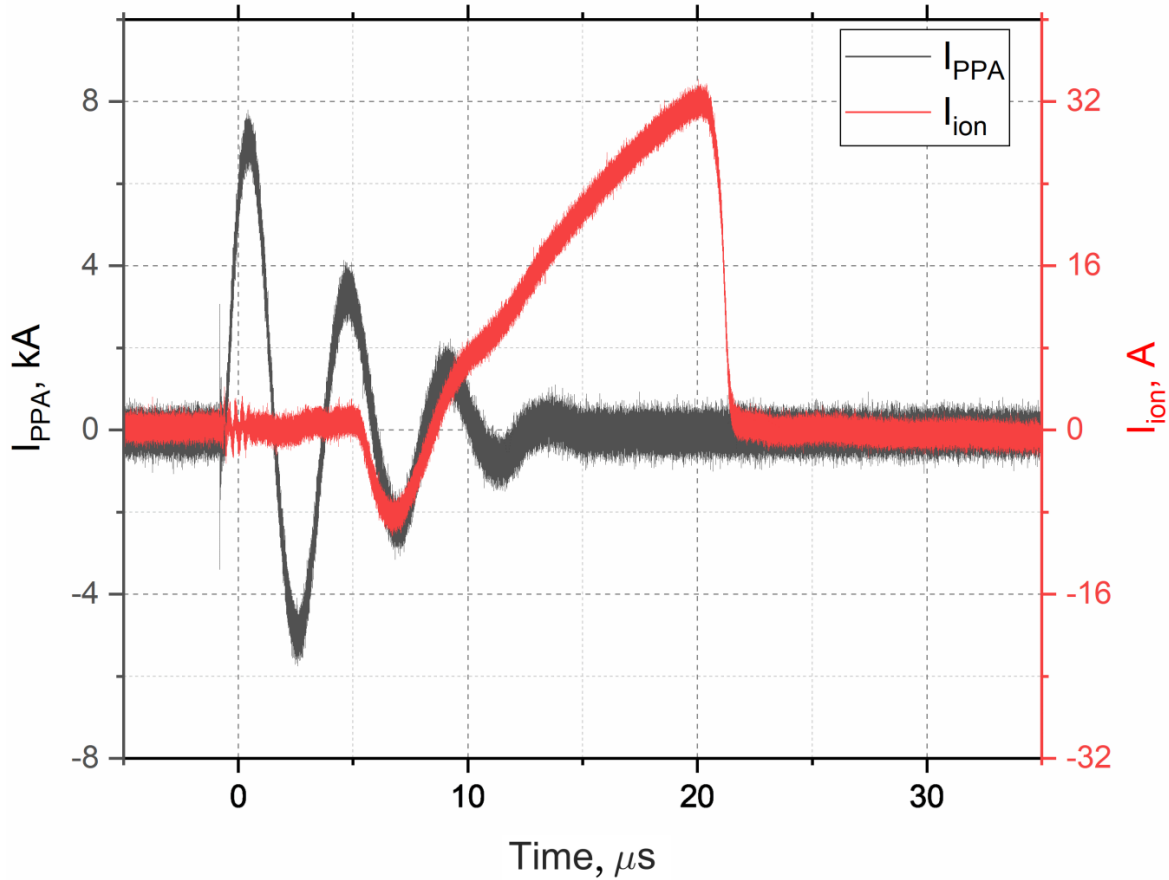

**Figure 6: LF-PPT discharge current and total exhaust plume current measured by the single Langmuir probe.**

**Conclusions**

A liquid-fed pulsed plasma thruster was presented in this work with pentaphenyl trimethyl trisiloxane propellant. An initial characterization of the thruster was conducted, including electrical parameter measurements of pulsed plasma accelerator and LESF igniter, and visual demonstration of the plasma dynamics. Time-of-flight measurements were used to estimate ion velocities in excess of 32 km/s. Thrust and impulse bit were estimated at 5.8 N and 35 μN·s, respectively, based on total ion current measurements. The results of this paper provide valuable information to enable development of a flight-ready LF-PPT. Propellant optimization, numerical simulation, longevity studies, and a comprehensive performance analysis are planned in ordinance with this development.

**Acknowledgements**

We would like to thank Xingxing Wang, Lee Organski, and Animesh Sharma for their help over the course of this research.